\documentclass[10pt,twoside]{article}
\usepackage{pkas2}
\usepackage{graphicx}
\makehead{David L Clements}{The AKARI Deep Field South}

\def\lesssim{\mathrel{\hbox{\rlap{\hbox{\lower4pt\hbox{$\sim$}}}\hbox{$<$}}}}
\def\gtrsim{\mathrel{\hbox{\rlap{\hbox{\lower4pt\hbox{$\sim$}}}\hbox{$>$}}}}

\begin{document}
\sloppy
\pagenumbering{arabic}
\twocolumn[
\pkastitle{27}{1}{2}{2012}
\begin{center}
{\large \bf {\sf
The AKARI Deep Field South:
A New Home for Multiwavelength Extagalactic Astronomy}}\vskip 0.5cm
{\sc David l Clements}\\
Astrophysics Group\\
Blackett Lab, Imperial College, Prince Consort Road, London SW7 2AZ, UK\\
{\it {E-mail: d.clements@imperial.ac.uk} }\\
\normalsize{\it (Received July 1, 2012; Accepted ????)}
\end{center}
\newabstract{
The importance of multiwavelength astronomical surveys is discussed in the context of galaxy evolution. The AKARI Deep Field South (ADF-S) is a new, well placed survey field that is already the subject of studies at a wide range of wavelengths. A number of ADF-S observational programmes are discussed and the prospects for the ADF-S as a future resource for extragalactic astronomy is explored.
\vskip 0.5cm
{\em key words:} infrared - galaxies: infrared - surveys}
\vskip 0.15cm  \flushbottom
]

\newsection{INTRODUCTION}

The availability of mutiwavelength surveys has become increasingly important in astrophysics. Where once there were specialists in optical, infrared or submillimetre astronomy, astronomers now study specific classes of object or physical problems, and bring to bear a wide range of observational techniques at a wide range of wavelengths to study their chosen targets. This is especially true of survey observations where the science goals can be quite diverse, and where there is considerable legacy value for future studies that might not be considered by the original observers. The selection of good fields for studies across the electromagnetic spectrum and the coordination of observations at a wide range of space and ground-based observatories is thus an increasingly important matter.

The prime reason for the importance of multiwavelength astronomy in the context of galaxy evolution is that the spectral energy distributions (SEDs) of galaxies cover a wide range of wavelengths, with different wavebands dominated by different processes. Four galaxy SEDs for objects covering a range of star formation rates are shown in Figure 1 (Planck Collaboration, 2011a). These SEDs include data from the optical, near-IR, mid-IR, far-IR and mm/submm from instruments on the ground and in space. Also shown are models of these SEDs. Different parts of the SED are dominated by different aspects of the galaxies. The UV is powered largely by young massive stars, the optical and near-IR by the old stellar population, the mid-IR by warm dust and polycyclic aromatic hydrocarbons (PAHs), and the far-IR/submm by dust enshrouded star formation. Radio emission is also associated with star-formation (eg. Helou, Soifer \& Rowan-Robinson, 1985), which AGN and late stages in stellar evolution provide significant emission in X-rays. To fully understand all the processes underway in an individual galaxy one must thus cover the whole electromagnetic spectrum.

The study of galaxy evolution is also inherently multiwavelength. Since the late 1990s and the discovery of the Cosmic Infrared Background (CIB; Puget et al., 1996; Fixsen et al., 1998) it has been apparent that roughly 50\% of energy generation throughout the history of the universe has been obscured by dust and re-emitted into the rest frame far-IR. Objects such as Arp220, 99\% of whose bolometric luminosity emerges in the far-IR, are rare in the local universe, but they must have been much more common at higher redshifts for the CIB to be generated. The first submm surveys with SCUBA (eg. Eales et al., 2000; Hughes et al., 1998; Smail, Ivison \& Blain, 1997) confirmed this result, providing the first detections of submm galaxies, while surveys with Herschel (eg. Clements et al., 2010; Oliver et al., 2010) are now finding many thousands of these objects, largely in fields already selected as prime multiwavelength survey locations. One of these fields is the AKARI Deep Field South.

\begin{figure}[!ht]
\resizebox{\hsize}{!}{\includegraphics{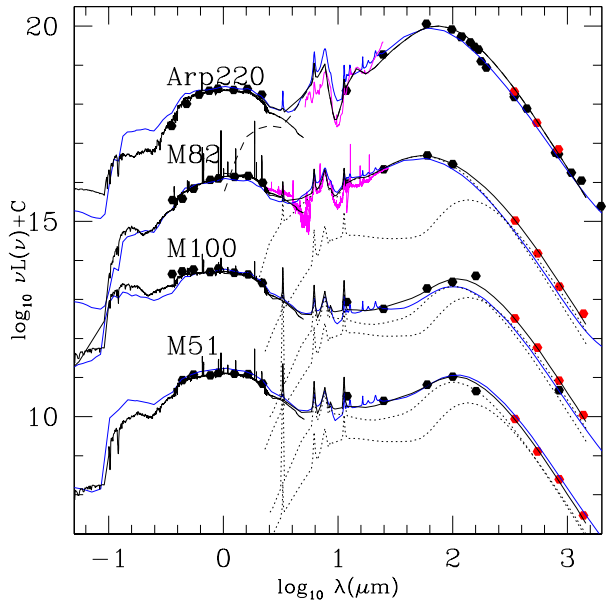}}
\caption{The spectral energy distributions of four example galaxies ranging from quiescent objects (M51 and M100) to more actively star-forming galaxies such as M82 and Arp220. Points are observed data points and curves are models. For more information see Planck Collaboration (2011a).}
\end{figure}

\newsection{THE AKARI DEEP FIELD SOUTH}

The AKARI Deep Field South (ADF-S; location given in Table 1) covers an area of $\sim$12 sq. deg. close to the south ecliptic pole, in a region of low cirrus and zodiacal dust foregrounds. It was observed by AKARI using both the Far-Infrared Surveyor (FIS; Kawada et al., 2007; Figure 2) at 65, 90, 140 and 160$\mu$m and the InfraRed Camera (IRC; Onaka et al., 2007) at 2.4, 3.2, 4.1, 7, 11, 15 and 24 $\mu$m. Numerous other facilities have also observed the ADF-S, a summary of which is available in Figure 3. In addition to these observations the field has also been observed by Herschel as part of the HerMES survey (Oliver et al., 2012). Its position close to the ecliptic pole also means that the field will be observed deeply by survey satellites that scan the sky using scans that are close to great circles in ecliptic coordinates. Examples of such missions include Planck, which is currently scanning the sky in nine bands at frequencies ranging from 857 to 30 GHz (Planck Collaboration, 2011b), and Euclid, which will provide deep near-IR imaging and spectroscopy across the sky after launch, currently scheduled for 2019 (Laureijs et al., 2010).

Scientific exploitation of the ADF-S is still in its early days, with many studies currently concentrating on specific wavebands while large scale mulitwavelength cross identification programmes await the public release of various catalogs and images. Nevertheless, the progress being made in these various studies demonstrates the growing database of sources in the ADF-S and promise much for the future.

In the remainder of this paper I will highlight results from various studies conducted in the ADF-S. The present volume also includes a variety of studies using data from this field.

\begin{figure*}[!ht]
\resizebox{\hsize}{!}{\includegraphics{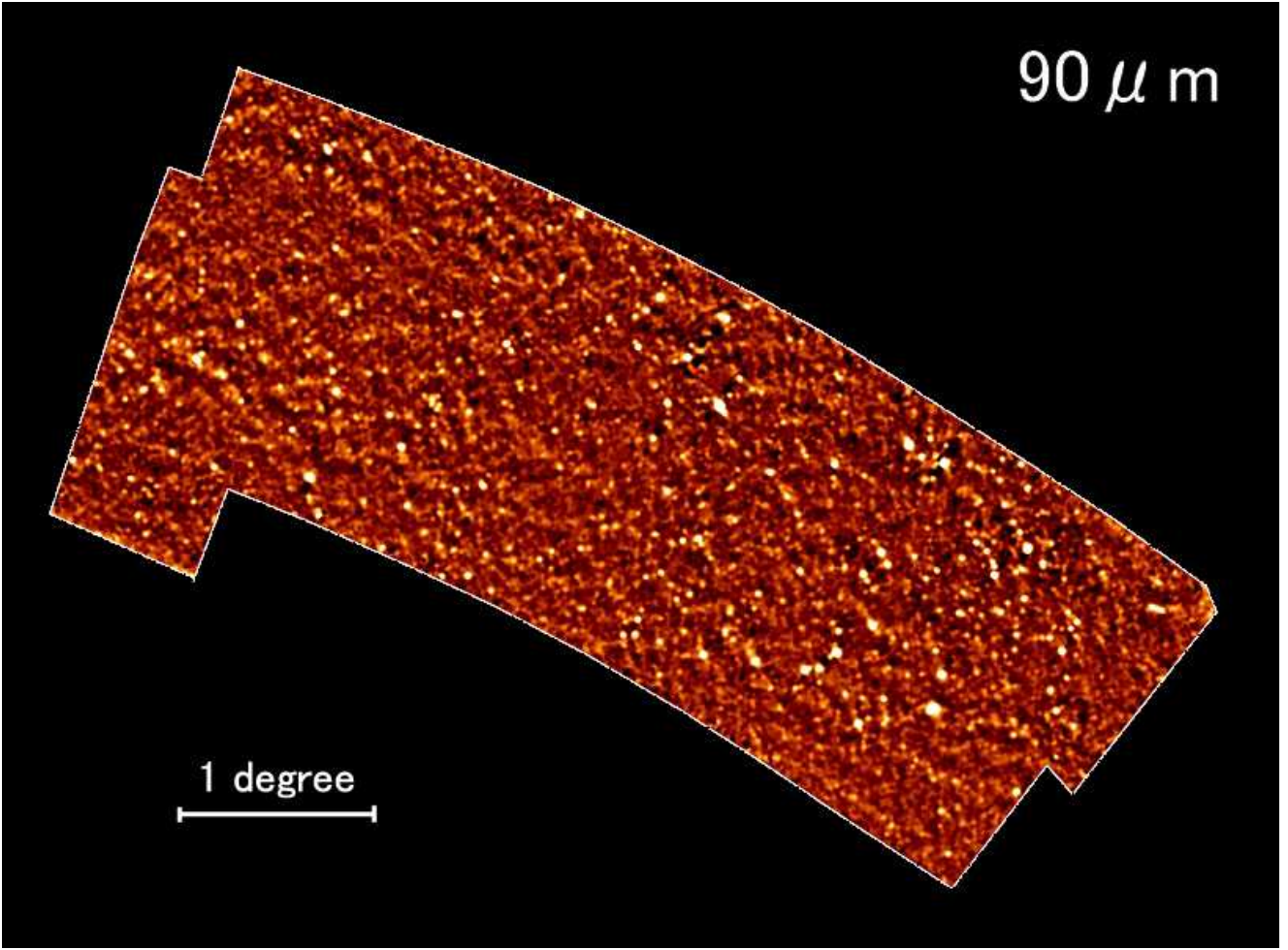}}
\caption{Image of the ADF-S at 90$\mu$m from the AKARI FIS.}
\end{figure*}

\begin{table}[!t]
\begin{center}
\scriptsize
\bf{\sc  Table 1.}\\
\sc{The Location of AKARI Deep Field South} \\
\begin{tabular}{cc}
\\ \hline \hline
RA&Dec\\ \hline
\hline
4:44:00&-53:20:00\\
\hline
\end{tabular}
\end{center}
\end{table}

\begin{figure*}[!ht]
\resizebox{\hsize}{!}{\includegraphics{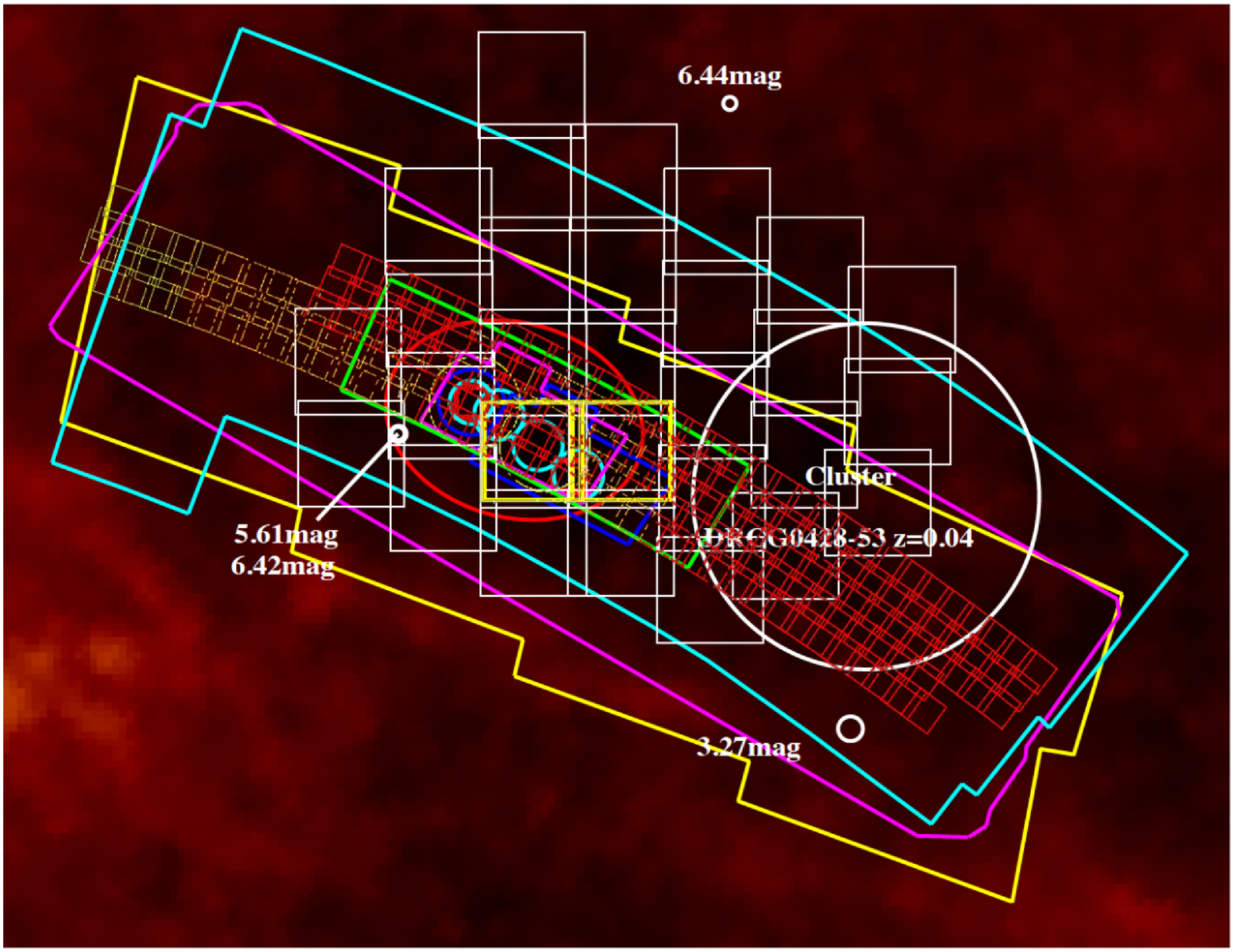}}
\resizebox{\hsize}{!}{\includegraphics{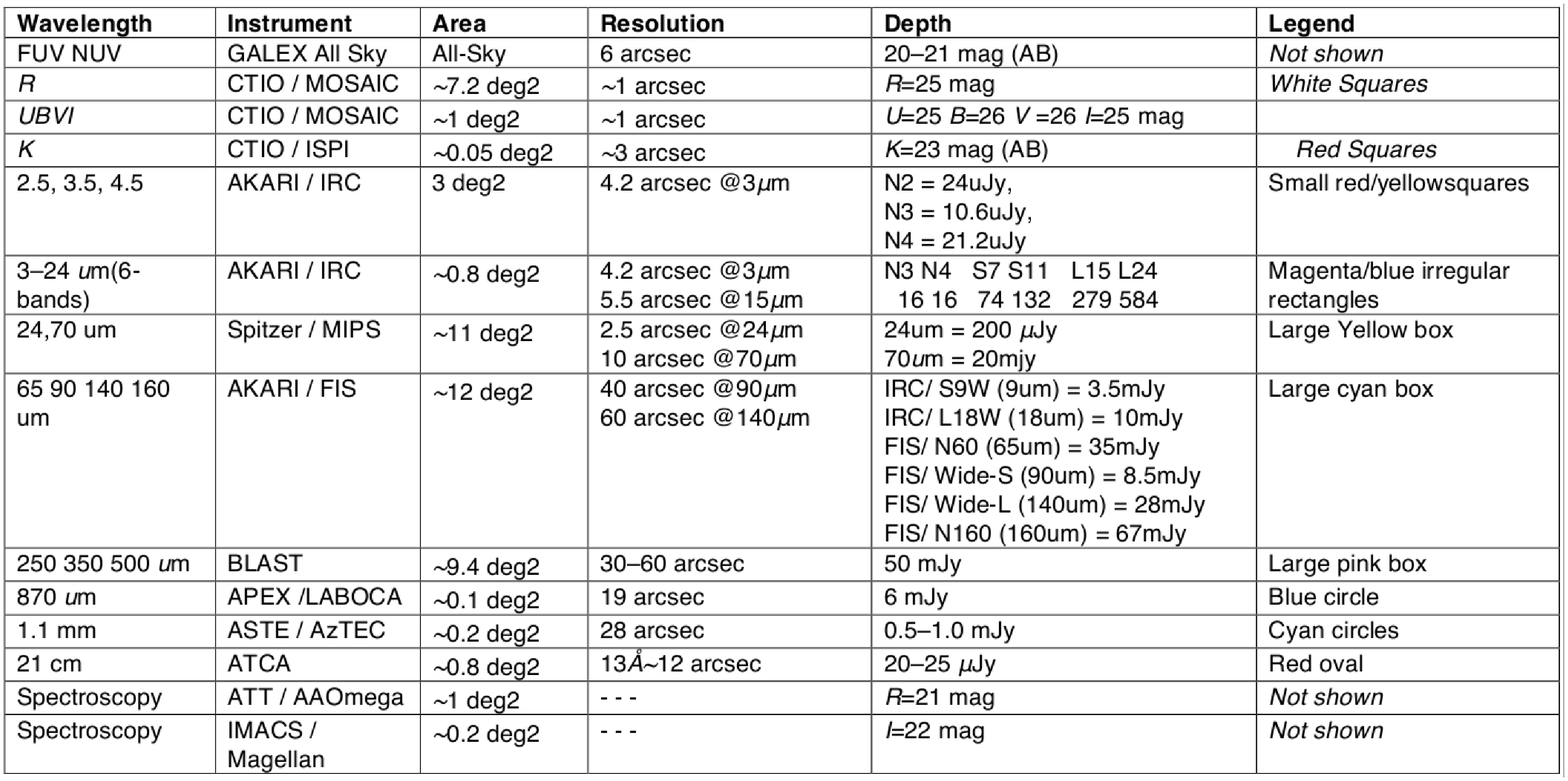}}
\caption{The various surveys in the ADF-S region overlaid on a map of the 100$\mu$m flux from IRAS and a summary of these observations compiled by C. Pearson.}
\end{figure*}

\newsection{NUMBER COUNT STUDIES}

Number counts are a fundamentally important tool in extragalactic astronomy and are often the first results to emerge from any new survey. Number count studies in the ADF-S have so far been published from the mid-IR to the radio. Using Spitzer data, Clements et al. (2011) produced counts at 24$\mu$m and 70$\mu$m (see also Scott et al., 2010). These counts are consistent with data obtained by other Spitzer surveys and with galaxy count models, though the presence of a foreground, z=0.04 galaxy cluster DC0428-53 (Dressler et al., 1980) somewhat complicates the interpretation.

At longer wavelengths the ADF-S was observed by the BLAST telescope (Valiante et al., 2010) at 250, 350 and 500$\mu$m, wavelengths that match those observed by the SPIRE camera on Herschel (Griffin et al., 2010). The BLAST counts are consistent with the counts later obtained by Herschel in these bands (Clements et al., 2010; Oliver et al., 2010; Glenn et al., 2010), though the specific Herschel observations of the ADF-S are still awaiting a full analysis.

At longer wavelengths still, Hatsukade et al (2011) conducted a survey at 1100$\mu$m of a 0.25 sq. deg. section at the centre of the ADF-S using the AzTEC instrument on the ASTE telescope. This is one of the largest area, deep (RMS of 0.3 - 0.7mJy) surveys in this band to date. Most of the 198 sources found by AzTEC lack counterparts in the AKARI bands, suggesting that they lie at z$>$1.5, a result confirmed by an initial cross-matching analysis with the Herschel catalogs in the ADF-S (Clements et al., in preparation). 

Counts for radio sources in the ADF-S have been produced by White et al. (2012) who used ATCA to survey 2.5 sq. degs. at 1.4 GHz to sensitivities of 19-50 $\mu$Jy/beam. They uncover 530 sources with counts that suggest that the majority of sources below 1mJy in flux are star-forming galaxies rather than AGN. Cross identification of these radio sources with Spitzer and AKARI data, to determine the colours of the radio sources, confirms this interpretation.

\newsection{CROSS IDENTIFICATIONS}

Counts in bands from mid-IR to submm are usually interpreted in the context of backward evolution models, whereby the present-day galaxy population is evolved backwards in time, to higher redshifts, using models that parameterise the various aspects of this evolution (e.g. Bethermin et al., 2012). The predictions of these models can then be tested against observations of redshift distributions, luminosity functions etc.. To be able to perform these tests, sources identified in the various surveys must be cross identified typically with optical counterparts so that spectroscopic observations can be used to obtain redshifts, and so that other properties of these sources can be determined. Cross-identification programmes in the ADF-S includes work with both existing multiwavelength source databases such as NED (Malek et al., 2010), and with optical imaging data that is then used as the basis of multiobject spectroscopy campaigns. Optical spectroscopy using AAOmega  (Sedgwick et al., 2011) has allowed the first luminosity function determination so far for AKARI 90$\mu$m sources in the ADF-S. The results are broadly consistent with existing evolution models, though there are some issues that may be attributed to large scale structure effects, since the sample size, at 389 galaxies, is currently rather small. Deeper optical imaging data has been obtained for the ADF-S, so the cross identification programmes will certainly expand in both size and scope, especially with the addition of the deeper, higher resolution Herschel data. 

\newsection{THE COSMIC INFRARED BACKGROUND}

As well as studies of individually detected objects, the AKARI maps can, after the removal of discrete sources and zodiacal dust, be used to determine the statistical properties of the CIB. This has been undertaken by Matsuura et al. (2011). They find that the CIB brightness is consistent with results from COBE, and that the CIB fluctuation spectrum is consistent with results from Spitzer and BLAST. They also find that the CIB fluctuations are red in colour, suggesting that ULIRG-like objects at high redshift are responsible for its production.

\newsection{DISCUSSION AND CONCLUSIONS}

The work on the ADF-S presented in this paper is still at an early stage. As has been noted, there has so far been little work published that uses cross identification of sources across the wide range of wavelengths where catalogs are currently available, but it is already clear that the ADF-S is proving to be a very useful resource. As the various different catalogs are stitched together through cross identifications, and as the complementary data at optical, near-IR and other wavelengths improves, the ADF-S will become increasingly important. In particular, its location as a well studied, low cirrus background field in the southern hemisphere will make it an ideal target for the next generation of telescopes being built in the southern hemisphere. VLT is already in full operation, and will surely be a key facility for the optical and near-IR followup of interesting sources in this field. ALMA will be of special importance for following up the dusty objects detected in the field by AKARI, Herschel and AzTEC, while future facilities such as E-ELT and CCAT will all have major roles to play.

The story of the ADF-S is only just beginning.

%

\acknowledgments{This paper would have been impossible without the invaluable help of Chris Pearson and all the members of the broad AKARI Deep Field South team. It is a privilege to be able to thank them here. This research has made use of the NASA/IPAC Extragalactic Database (NED) which is operated by the Jet Propulsion Laboratory, California Institute of Technology, under contract with the National Aeronautics and Space Administration and is supported in part by the UK STFC.
}

\end{description}

\end{document}